\documentclass{blois}

\usepackage{cite}
\def\be{\begin{equation}}
\def\ee{\end{equation}}
\def\bea{\begin{eqnarray}}
\def\eea{\end{eqnarray}}

\newcommand{\Photo}{}

\begin{document}
\vspace*{4cm}
\title{SHiP: A NEW FACILITY TO SEARCH FOR LONG LIVED NEUTRAL PARTICLES AND INVESTIGATE THE $\nu_\tau$ PROPERTIES\\Proceedings of the 28\textsuperscript{th} Rencontres de Blois, 2016}

\author{ ELENA GRAVERINI on behalf of the SHiP Collaboration }

\address{Physik Institut der Universit\"at Z\"urich\\Winterthurerstrasse 190,
8057 Z\"urich, Switzerland}

\maketitle\abstracts{
SHIP is a new general purpose fixed target facility, whose Technical Proposal has been recently reviewed by the CERN SPS Committee and by the CERN Research Board. The two boards recommended that the experiment proceeds further to a Comprehensive Design phase. A 400 GeV proton beam extracted from the SPS will be dumped on a heavy target with the aim of integrating $2\times 10^{20}$ proton-target collisions (\textit{pot}) in 5 years. A dedicated detector, based on a long vacuum tank followed by a spectrometer and particle identification detectors, will allow to probe a variety of New Physics models with light long-lived exotic particles and masses below $\mathcal{O}(10)$~GeV/$c^2$, including Dark Photons, light scalars and pseudo-scalars, and Heavy Neutrinos. The sensitivity to Heavy Neutrinos will allow for the first time to probe, in the mass range between the $K$ and the $D$ meson mass, a coupling range for which Baryogenesis and neutrino oscillations could also be explained. Another dedicated detector will allow the study of neutrino cross-sections and angular distributions. $\nu_\tau$ deep inelastic scattering cross sections will be measured with a statistics 1000 times larger than currently available, with the extraction of the so far never measured $F_4$ and $F_5$ structure functions, and allow to perform charm physics studies with significantly improved accuracy.
}

\section{Introduction and physics motivation}
The Standard Model (SM) consistently describes all known microscopic physics phenomena. However, the observed neutrino oscillations, the existence of Dark Matter and the matter-antimatter asymmetry in the Universe provide established experimental evidence that prevents the SM from being seen as the ultimate theory of Nature.
Dark Matter can be found naturally in New Physics (NP) models providing a so-called ``Hidden Sector'' (HS).
An example is the $\nu$MSM model~\cite{nuMSM}, a minimalistic extension that adds three right-handed neutrinos $N_{1,2,3}$, also called sterile or Majorana neutrinos or Heavy Neutral Leptons (HNLs), to the particle content of the Standard Model.
One of the new states, $N_1$, has a mass in the keV region and a lifetime exceeding that of the Universe, therefore providing a Dark Matter candidate. The other two states $N_{2,3}$ both have mass in the MeV-GeV range, being almost degenerate, and are responsible for the masses of the active neutrinos through the Seesaw mechanism, as well as providing an explanation for the observed matter-antimatter asymmetry. The latter can be generated through leptogenesis thanks to the Majorana mass term that can accompany the heavy neutrino states, which are $SU(2)$ singlets.

The proposed SHiP experiment provides a general-purpose fixed target facility using the SPS accelerator complex.
The project aims at investigating the presence of very weakly interacting, long lived particles that do not interact directly with the SM, and that are coupled to it through gauge-singlet operators (``portals''). In the context of $\nu$MSM sterile neutrinos, SHiP would allow to probe most of the allowed parameter space below a few GeV.

In addition to the New Physics program, a beam dump facility like SHiP allows for abundant production of neutrinos of all species in charmed hadron decays. Therefore, SHiP will include a dedicated detector to study the properties and cross sections of the $\tau$ (anti)neutrinos, allowing for the first observation of the $\bar{\nu}_\tau$ and the measurement of the unknown $F_4$ and $F_5$ nucleon structure functions.
A detailed description of the SHiP physics case can be found in~\cite{PP}.

\section{The SHiP experiment}
\begin{figure}
	\centering
	\raisebox{-0.5\height}{\includegraphics[width=0.37\textwidth]{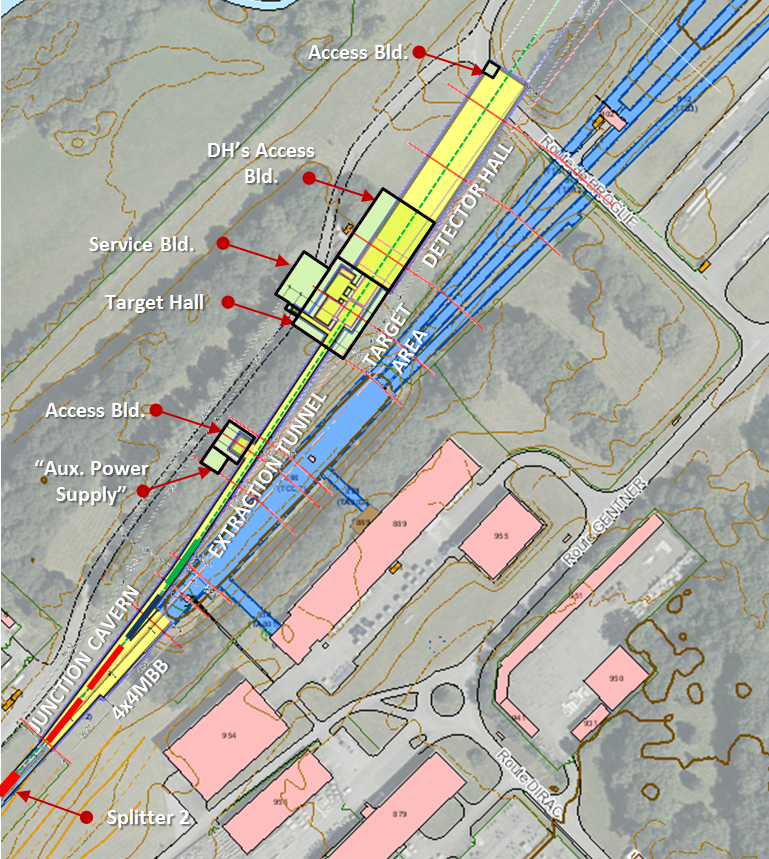}}
	\raisebox{-0.5\height}{\includegraphics[width=0.62\textwidth]{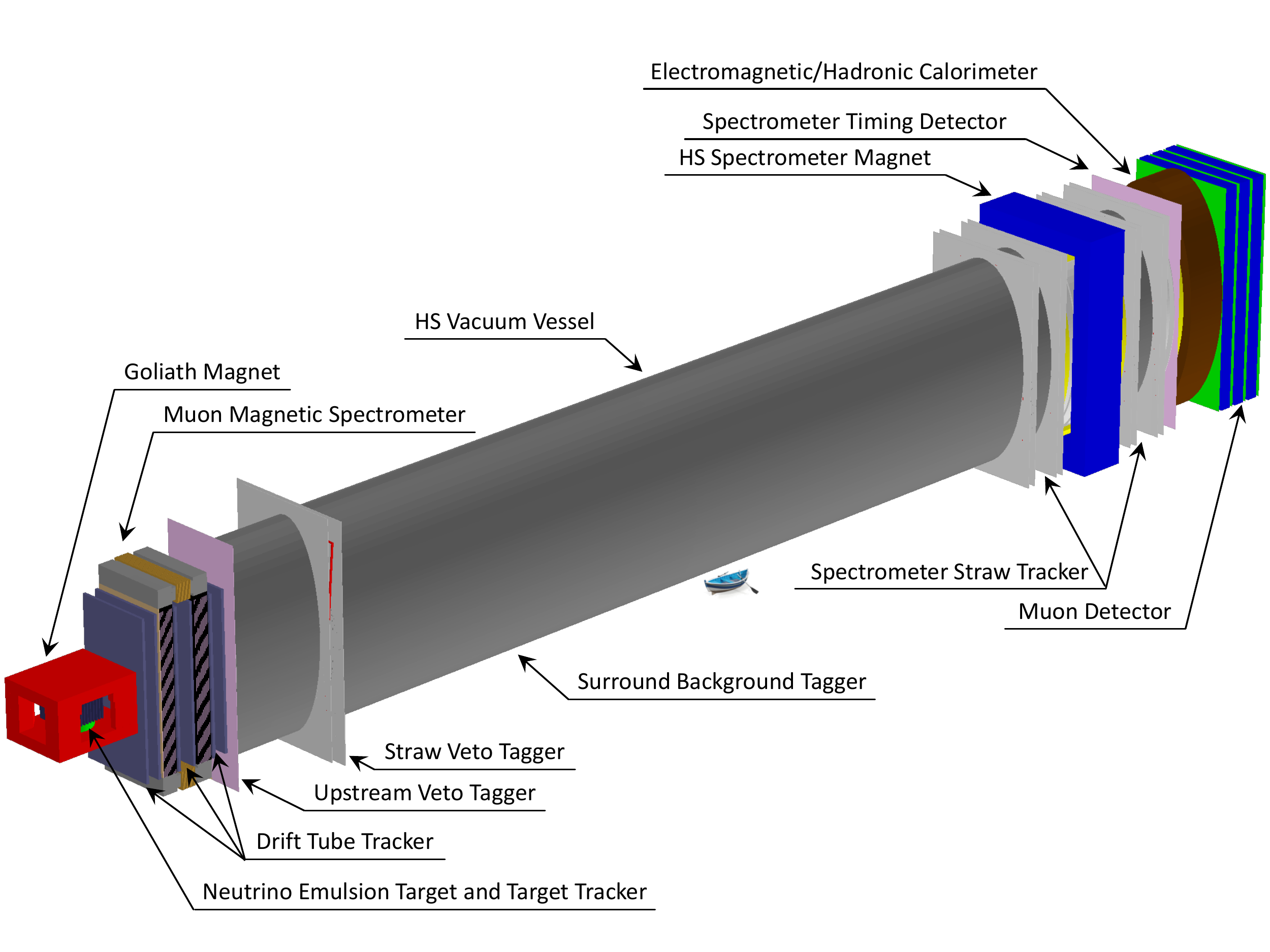}}
	\caption{Left: overview ofthe SHiP facility. Right: detailed view of the SHiP tau neutrino and hidden sector detectors (\textsc{Geant4}).}\label{img:hsdetector}
\end{figure}

SHiP will search for hidden particles produced in decays of charmed mesons and decaying to charged SM particles.
A dedicated beam line will be branched off the SPS extraction line at the CERN North Area. The proposed location of the target and experimental hall~(Figure~\ref{img:hsdetector}) allows for future extensions of the detector complex.
The 400 GeV/$c$ proton beam will be delivered onto a heavy target, made of alternated blocks of titanium-zirconium doped molybdenum and tungsten, with a centre-of-mass energy $E_{CM} = \sqrt{2\, E_{b}\, m_p} \simeq 27$~GeV.
Under nominal running conditions, the SPS will deliver $2 \times 10^{20}$ protons on the SHiP target in 5 years of operation~\cite{TP}.
The target is followed by a hadron absorber made of iron, with the aim of stopping the hadronic and electromagnetic products and any residual non-interacting protons.
A very large fraction of the muons emerging from the beam dump will be deflected outside of the SHiP fiducial volume by means of a compact active shielding system based on magnets of alternate polarity, which design was optimised in order to cope with the expected muon spectrum obtained from simulation.

A dedicated neutrino detector will be placed immediately upstream of the Hidden Sector detector, taking advantage of the muon-cleared area. 
It will consist of a magnetised target made of bricks of alternated layers of laminated lead and nuclear emulsion foils, similar to those used by the OPERA experiment~\cite{opera}, followed by a target tracker and by a muon magnetic spectrometer. This layout has been already extensively tested at OPERA and it has been shown to be able to detect all three flavours of neutrinos efficiently. The target magnetisation, together with the use of Compact Emulsion Spectrometers, will allow to measure the charge of the produced hadrons and therefore to separate $\nu_\tau$ and $\bar{\nu}_\tau$.
Details about the $\nu_\tau$ detector and its physics case are given in~\cite{TP, PP}.

The detector for hidden particles is designed to fully reconstruct their decays into charged SM particles and to reject the background due to SM processes down to less than 0.1 events in 5 years of data taking.
The fiducial volume is contained in a 62~m long vacuum vessel with elliptical cross-section of $5 \times 10$~m$^2$. A level of vacuum of $10^{-6}$~atm will be needed to reduce the background arising from neutrino interactions in air to less than one event in 5 years. 

The length of the fiducial volume is 50~m, and it is followed by a 12~m long magnetic spectrometer.
The latter will consist of four stations symmetrically arranged around a large aperture dipole magnet providing an integrated field of 0.65~Tm. Each of the four stations consists of two stereo and two $y$ layers of 5~m long straw tubes.

Outside of the vacuum vessel, a high accuracy timing detector will be placed. A particle identification system will follow, featuring electromagnetic and hadronic calorimeters (both using the shashlik technology) followed by a muon system made of four active layers interlaced with iron.
An upstream tagger will help to detect and veto charged particles produced in front of the main decay volume. A straw tagger is placed in vacuum 5~m downstream of the entrance lid of the vessel. An additional background tagger surrounds the fiducial decay volume, which walls enclose 30~cm of liquid scintillator.

\subsection{Background sources and strategies}

Thanks to the intense production of heavy hadrons in its target, SHiP will greatly improve the sensitivities of previous experiments on models with a secluded sector.

The main background to the hidden particle decay signal originates from the inelastic scattering of neutrinos and muons upstream of the detector or in its vicinity, producing long-lived neutral mesons.
Another source of background are random combinations of tracks from the residual muon flux, or other charged particles from inelastic interactions in the proximity, which enter the decay volume and together mimic signal events.
Cosmic muons can contribute to both types of background, but their yield is expected to be small~\cite{TP}.
Table~\ref{tab:background} shows the level of rejection of the main backgrounds that is attainable at SHiP, according to detailed simulations~\cite{TP}.

\begin{table}
	\centering
	\caption{\small Background yields from simulation for different background sources.
		In all cases, zero events remain after applying all the selection criteria. 
		Hence, an upper limits at 90 \% CL is calculated as UL(90\%) = $-\ln (0.1)/{\rm weight}$, where the weight is the ratio between the generated sample and the expected yield for $N_{\rm pot} = 2 \cdot 10^{20}$.
		For the muon inelastic background, the upper limit is conservatively calculated by ignoring the
		factorizability of the veto efficiencies for the incoming $\mu$ and for the particles produced 
		in the muon interaction. Assuming factorizability, the yield is reduced down by another factor $10^{-3}$~\cite{TP, addendum}.}
	\label{tab:background}
	\begin{small}
		\begin{tabular}{|lcc|}
			\hline
			Background source & Strategy & Yield (UL 90\% CL) \\ \hline
			{\bf $\nu$-induced} & upstream and surround taggers, topology & 3.3\\
			
			{\bf $\overline{\nu}$-induced} & upstream and surround taggers, topology  & 2.1\\
			
			{\bf Muon inelastic} & upstream and surround taggers, topology &  4.6 \\ 
			{\bf Muon combinatorial} & timing detector & 0.1 \\ 
			
			{\bf Cosmics} & surround tagger, topology & 1.2\\\hline
		\end{tabular}
	\end{small}
	\vspace{-3mm}
\end{table}

Accurate simulation studies indicate that a level of background of 0.1 events for 5 years of data taking is achievable, thanks to the redundant system of veto detectors.
SM neutrino interactions producing a hidden particle candidate were found to take place mainly in the muon magnetic spectrometer of the tau neutrino detector, and in the entrance window and the surrounding walls of the vacuum vessel. They are therefore efficiently vetoed by the upstream and surround taggers, and by analysing the event topology.
Neutrino interactions with the residual gas in the decay volume are negligible if the vacuum pressure is below $10^{-6}$ bar.
Hidden particle candidates arising from muon inelastic scattering, consisting mainly of $K_L$, $K_S$, $\Lambda$ and $\pi^\pm$ decays and $\gamma$ conversion, are also efficiently tagged by the upstream vetoes and by the surround liquid scintillator tagger.
Combinatorial muons are randomly distributed over the duration of each spill: this kind background is fought with the use of timing requirements.
Finally, the topology of events due to cosmic rays is such that simple selection criteria based on the event topology reduce this background to a negligible level.

\subsection{Expected sensitivity}
\begin{figure}
	\centering
	\raisebox{-0.55\height}{\includegraphics[width=0.5\textwidth]{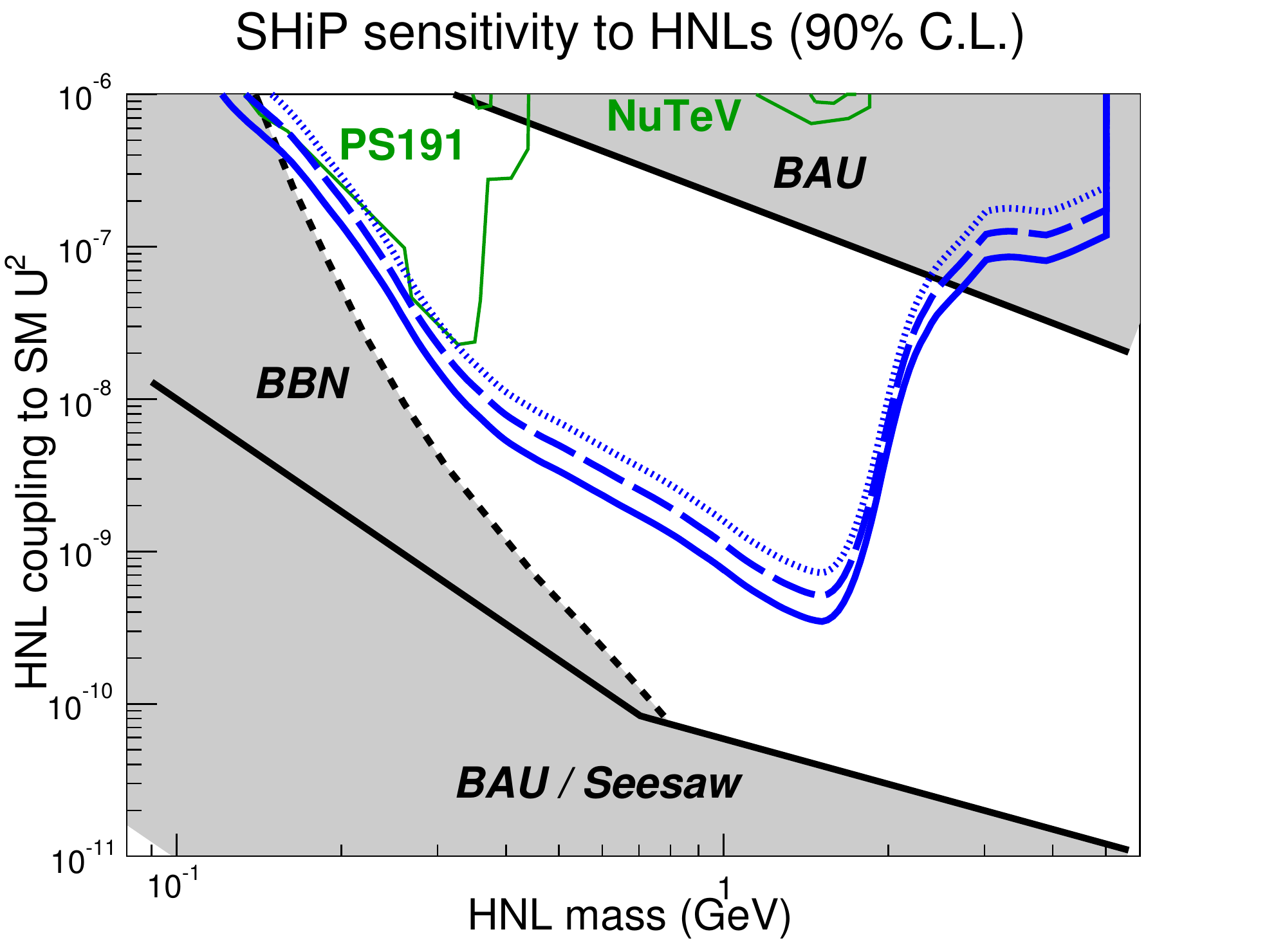}}\hfill
	\raisebox{-0.55\height}{\includegraphics[width=0.5\textwidth]{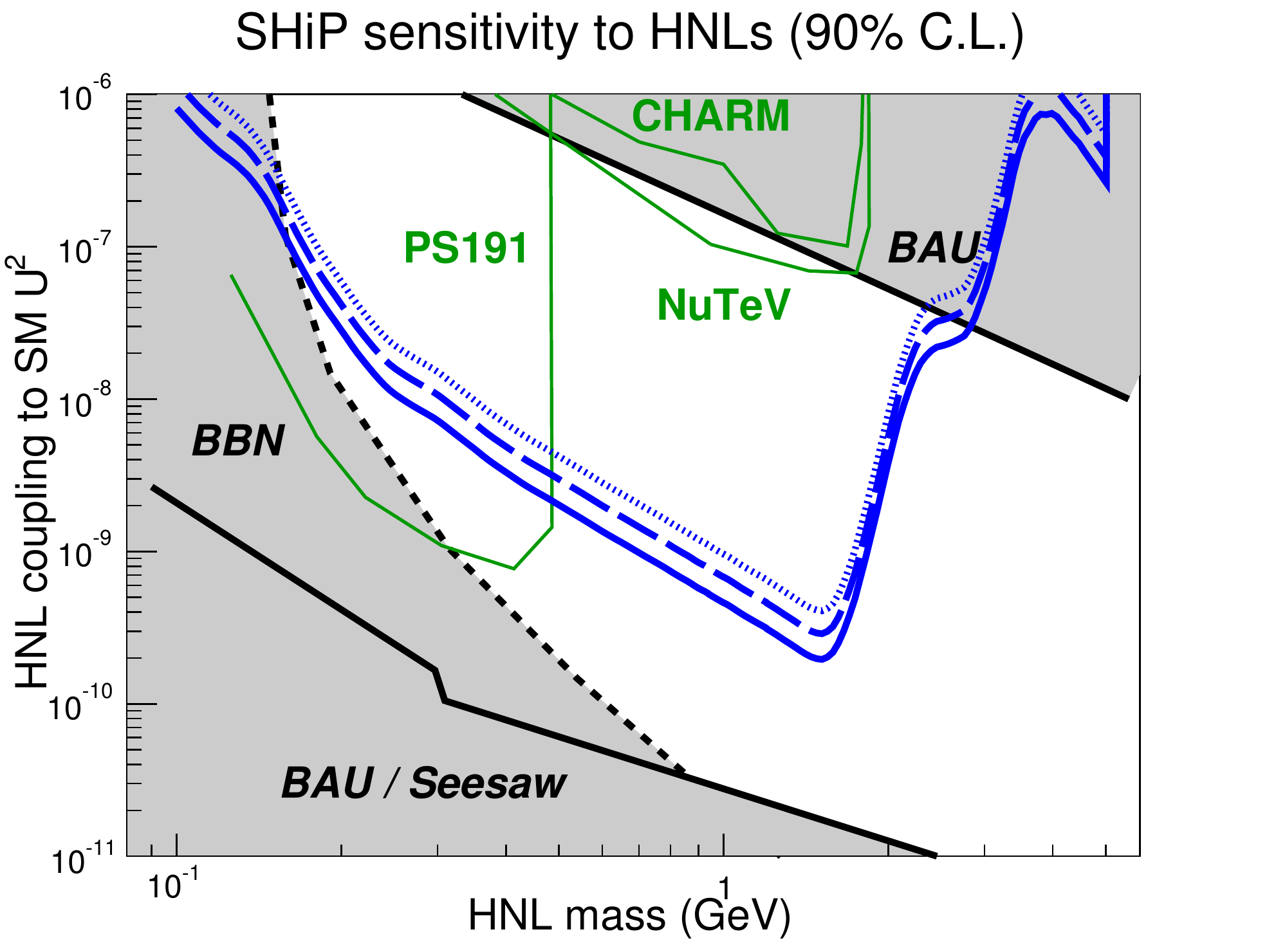}}
	\caption{SHiP's discovery potential in the parameter space of the $\nu$MSM, for a level of background of 0.1 (continuous line), of 10 (dashed line), and of 10$\pm$6 (dotted line) events in 5 years. Left: normal hierarchy, $U_\mu^2$ dominating; right: inverted hierarchy, $U_e^2$ dominating~\cite{Graverini:2214085}.}\label{img:hnl-sens}
\end{figure}

An extensive review of the sensitivities that SHiP can reach in all the probed models was published in~\cite{PP}. Here the sensitivity to HNLs is shown as an example.
In 5 years of data taking, the SHiP experiment can achieve sensitivities which are up to four orders of magnitude better than previous searches. In the context of the $\nu$MSM, SHiP can explore most of the cosmologically interesting region of the parameter space below the mass of the $D$ meson, reaching down to couplings close to the minimum values able to explain neutrino oscillations. HNLs with very low mass and couplings are excluded by observation of the relative abundance of hydrogen and heavier elements in the Universe (Big Bang Nucleosynthesis). Also very large couplings to the SM flavours are excluded, because they would not explain the observed matter-antimatter asymmetry. Figure~\ref{img:hnl-sens} shows SHiP's sensitivity to HNLs in two different scenarios, assuming normal and inverted hierarchy of the SM neutrino masses, and where the coupling to the muon flavour and to the electron flavour dominate, respectively.

About 1800 $\nu_\tau$ and 900 $\bar{\nu}_\tau$ interactions will be detected during the SHiP run. This will allow to measure the $F_4$ and $F_5$ structure functions, which are negligible in $\nu_\mu$ and $\nu_e$ interactions. The neutrino detector will also allow to probe the strange-quark content of the nucleon, thanks to the abundant production of charmed hadrons expected in anti-neutrino interactions.

\section*{References}

\bibliography{bibliography}

\end{document}